\documentclass[twocolumn,prl,amsmath,amssymb,raggedbottom,super]{revtex4}

\usepackage{graphicx}
\usepackage{dcolumn}
\usepackage{bm}
\usepackage{natbib}

\begin{document}
\title{A nuclear magnetic resonance spectrometer for operation around 1 MHz with~a~sub~10~mK
noise temperature, based on a two-stage dc SQUID sensor}%
\author{L.V.~Levitin}%
\author{R.G.~Bennett}
\author{A.~Casey}%
\email{a.casey@rhul.ac.uk}{}%
\author{B.P.~Cowan}%
\author{C.P.~Lusher}%
\email{c.lusher@rhul.ac.uk}%
\author{J.~Saunders}%
\affiliation{Department of Physics, Royal Holloway University of London, Egham, Surrey, TW20 0EX, UK}%
\author{D.~Drung}%
\author{Th.~Schurig}%
\affiliation{Physikalisch-Technische Bundesanstalt, Abbestrasse
2-12, D-10587 Berlin, Germany}%
\date{\today}

\begin{abstract}
We have developed a nuclear magnetic resonance  spectrometer with a
series tuned input circuit for measurements on samples at
millikelvin temperatures based on an integrated two-stage
superconducting quantum interference device current sensor, with an
energy sensitivity $\varepsilon = 26\pm1~h$ when operated at 1.4~K.
To maximize the sensitivity both the NMR pickup coil and tuning
capacitor need to be cooled, and the tank circuit parameters should
be chosen to equalize the contributions from circulating current
noise and voltage noise in the SQUID. A noise temperature $T_{N} =
7\pm2~\text{mK}$ was measured, at a frequency of 0.884 MHz, with the
circuit parameters close to optimum.

\end{abstract}

\maketitle

The extremely high sensitivity of superconducting quantum
interference devices (SQUIDs) to magnetic flux has been exploited in
nuclear magnetic resonance (NMR) spectrometers by a number of groups
for measurements on both cryogenic and room temperature
samples.\cite{Greenberg} In most of these applications the input
circuit is broadband with the NMR pickup coil and SQUID input coil
forming a superconducting flux transformer, however one can obtain
an improved signal-to-noise ratio ($S/N$) with SQUIDs, at the
expense of bandwidth, by tuning the input circuit. Clarke \textit{et
al.}\cite{CTG,Martinis} considered noise mechanisms and optimization
of dc SQUID circuits for tuned inputs. SQUID amplifiers with tuned
input circuits have been used for nuclear quadrupole
resonance,\cite{Hilbert1} NMR\cite{Freeman, Casey} and magnetic
resonance imaging.\cite{Seton}

Cooled preamplifiers have been employed to reduce noise in NMR
systems with tuned source impedances. Performance can be
characterized by the noise temperature $T_{N}$ (the temperature of
the source at which its contribution to the total output noise power
equals that of the amplifier). For a SQUID amplifier with a tuned
input the optimum $T_N$ is proportional to frequency. For our
experiments on NMR samples at millikelvin temperatures we use
frequencies $\sim$1~MHz, at which we may obtain a good $S/N$ with
acceptable rf heating. Cooled GaAs MESFETs have been used to detect
NMR signals at these frequencies, achieving $T_{N}$ $\sim$ 1~K.
\cite{Richards} Using a dc SQUID in open loop mode Freeman
\textit{et al.}\cite{Freeman} obtained a $T_N$ of 300~mK at 1.9 MHz,
limited by the readout electronics. Improved noise performance is
possible using two-stage SQUID amplification, and this provides the
primary motivation for this work. Two-stage SQUID amplifiers have
been used by Falferi \textit{et al.}\cite{Falferi2} in a resonant
gravitational wave detector, with an estimated $T_{N}$ of 15$~\mu$K
at 11~kHz, and by M\"{u}ck \textit{et al.},\cite{Muck} who obtained
a $T_{N}$ around 50~mK at 0.5~GHz with a microstrip SQUID cooled to
20~mK.

In this work we have developed a tuned NMR spectrometer based on an
integrated two-stage low-transition-temperature (low $T_c$) dc SQUID
sensor. The spectrometer sensitivity is limited by Johnson noise in
the resistive element of the input circuit and by SQUID amplifier
noise. The goal of this work was to study the limit imposed by the
SQUID noise. To that end we measured the spectrometer noise as a
function of the temperature of the resistive pickup coil, in the
absence of a sample. The pickup coil was placed in a superconducting
shield to suppress external interference. Performance of a similar
SQUID with a broadband superconducting input circuit used for NMR is
reported elsewhere.\cite{Casey2}

\begin{figure}[t]
\includegraphics{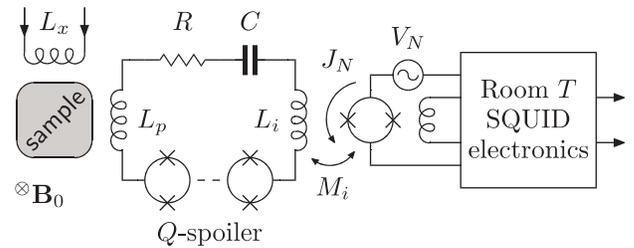}
\caption{\label{fig:setup} Schematic diagram of the tuned NMR
spectrometer. The two-stage SQUID sensor is depicted here as a
single SQUID. The device contains an integrated input coil $L_{i}$ and $Q$-spoiler. The temperature of
the pickup coil $L_{p}$ and capacitor $C$ can be varied. Inductance
$L_{x}$ represents the NMR transmitter coil.}
\end{figure}

A schematic diagram of the tuned NMR spectrometer is shown in
Fig.~\ref{fig:setup}. The pickup coil $L_{p}$ forms a series
resonant circuit with a capacitor $C$ and the SQUID input coil
inductance $L_{i}$. $R$ is the resistance in the input circuit. In
our case $R$ is mainly associated with the copper pickup coil, but
has contributions from dissipation in the capacitor and from the
presence of the SQUID. The SQUID sensor consists of a single SQUID
first stage, read out by a 16-SQUID series array, integrated onto a
single chip.\cite{Drung} The SQUID, mounted at 1.4~K (rather than
4.2~K) for improved noise performance, is connected directly to the
room temperature readout electronics.\cite{XXF} This permits
flux-locked loop (FLL) operation without flux modulation, and large
FLL bandwidths of dc up to 6~MHz. The device contains an array of 16
unshunted SQUIDs in the input circuit, which operates as a
$Q$-spoiler.\cite{Hilbert1,Freeman} This reduces the tank circuit
quality factor $Q$ for high signal levels and hence shortens the
recovery time from large current transients following removal of an
NMR transmitter pulse.

The experiment was mounted on a nuclear adiabatic demagnetization
cryostat, capable of achieving $\sim$ 200~$\mu$K, previously used
for our earlier NMR measurements on $^{3}\text{He}$ at low
millikelvin temperatures using a SQUID spectrometer with a tuned
input.\cite{Casey} The NMR pickup coil was wound from copper wire
and heat sunk to a plate whose temperature could be varied from
10 to 1500~mK. Care was taken to avoid thermal noise from dissipative
elements in the tuning capacitor by keeping it close to the coil
temperature.

Our two-stage SQUID is designed to have a single-SQUID-like
flux-voltage characteristic,\cite{Drung} with a large overall flux
to voltage transfer function $V_{\Phi}$. As shown in
Fig.~\ref{fig:setup}, we model the device as a single SQUID of
inductance $L_{s}$ (equal to that of the first stage SQUID), with a
voltage noise $V_N$ at the output, and a circulating current noise
$J_N$ in the SQUID loop. $V_N$ includes noise from the room
temperature amplifier as well as from the two SQUID stages. The
spectral densities of these noise sources, $S_V$, $S_J$ and a
correlation $S_{VJ}$, can be considered white above a few tens of
Hz.  In this work we write $S_V$ and $S_J$ in terms of a flux noise
$S_{\Phi}$ in the SQUID as follows:
\begin{equation}
S_{V} =V_{\Phi}^{2}S_{\Phi}^{\phantom{1}}\qquad \textrm{and}\qquad
S_{J} =\zeta^{2}S_{\Phi}^{\phantom{1}}/{L_s^2},\label{eq:spec den}
\end{equation}
where $S_{\Phi}$ is the total effective flux noise measured with the
input circuit open, and $\zeta$ is a dimensionless parameter. We
define the overall energy sensitivity as $\varepsilon =
S_{\Phi}/(2L_{s})$. Tesche \textit{et al.}\cite{Tesche} calculated
intrinsic noise in a single SQUID. They found $S_{VJ}$ to be real,
and $\zeta$ to be of the order of unity but dependent on SQUID bias.
We assume this to be the case for our two-stage device, since noise
from the first stage is dominant. A real $S_{VJ}$ does not
contribute to noise at the tank circuit resonance frequency.
\cite{Seton}

In an NMR experiment, precession of the magnetization induces a
signal voltage $V_S$ in the pickup coil. In this case the total flux
in the SQUID is
\begin{equation}
\phi = \left(V_S + V_{NR} - j\omega M_i J_N \right){M_i}/{Z_T} +
{V_N}/{V_{\Phi}}.\label{eq:total flux}
\end{equation}
Here $M_i$ is the mutual inductance between the input coil and the
SQUID, $Z_T = R+j(\omega L_{T} -1/\omega C)$ is the total impedance
in the input circuit, where $L_{T}= L_i+L_p$. $Z_T$ can be
influenced by coupling to the SQUID, mainly through a contribution
to $R$, which we neglect in this model. Changes in the SQUID
parameters due to strong coupling to the input circuit are also
neglected, since they are insignificant for the case of a high-$Q$
tuned source close to optimum.\cite{Hilbert2} $V_{NR}$ represents
the Johnson noise voltage in $R$, which is at temperature $T$. This
noise source is also white at the frequencies of interest. On
resonance the total flux noise is given by
\begin{equation}
S_{\Phi}^{\mathrm{(res)}} =\frac{4k_B^{\phantom{1}} TM_i^{2}}{R}+
S_{\Phi}\left[1+{\frac{\omega_0^2M_i^4{\zeta}^2}{L_s^2
{R}^2}}\right].\label{eq:fluxnoise}
\end{equation}
Let the signal $V_S$ be close to the tank circuit resonance
frequency $\omega_0 = (L_TC)^{-1/2}$ and measured in a bandwidth
$\Delta f$. The signal-to-noise ratio can be defined as
\begin{equation}\label{eq:SN}
S/N = V_S/{\sqrt{4k_B (T + T_N) R \Delta f}},
\end{equation}
where from Eq. \eqref{eq:fluxnoise}
\begin{equation}
T_N = \frac{1}{4k_B}\left[\frac{R}{M_i^2}+\frac{(\omega_0 M_i
\zeta)^2}{L_s^2 R}\right] S_\Phi.\label{eq:TN}
\end{equation}

For a given experiment the sample geometry and resonant frequency
$\omega_0$ are fixed. The choice of quality factor
$Q=\omega_{0}L_{\textrm{T}}/R$ is limited by the desired bandwidth
and the requirement of a sufficiently short recovery time following
a transmitter pulse. In order to optimize the sensitivity $L_p$, $R$
and $C$ can be varied simultaneously within these constraints.
Changing the number of turns in the pickup coil varies $V_S$ and
$L_p$ such that $V_S \propto \sqrt{L_p}$. Then Eq.~\eqref{eq:SN} can
be written as
\begin{equation}
(S/N)^2 \propto \frac{L_p}{4k_B (T + T_N) R \Delta f} =
\frac{Q(L_p/L_T)}{4k_B (T + T_N) \omega_0 \Delta f}. \label{eq:SNR}
\end{equation}
The highest $S/N$ is achieved by making $Q$ as large as possible
within the constraints. For $Q \gg 1$, $L_p/L_T$ is close to 1 when
$T_N$ is minimized (see Eq.~\eqref{eq:five}). Therefore the $S/N$ is
maximum when $T_N$ is minimum.

\begin{figure}[t]
\includegraphics{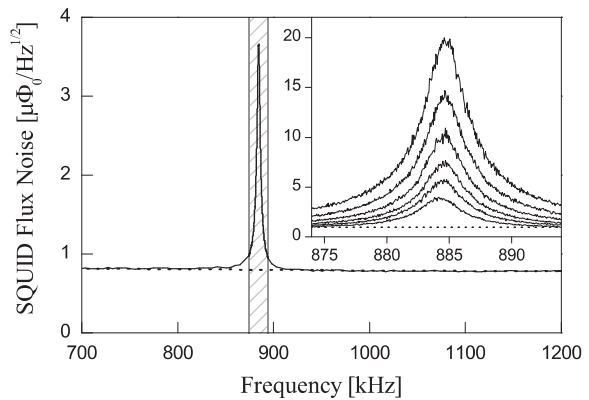}
\caption{\label{fig:flux_noise} Frequency dependence of the flux
noise referred to the first stage SQUID measured in open loop mode,
with the pickup coil at 20~mK. Off-resonance noise corresponds to
$\varepsilon = 26~{h}$. Both Johnson noise in the pickup coil and
circulating current noise in the SQUID contribute to the peak noise.
Inset shows the noise in the vicinity of the peak for coil
temperatures of 20, 55, 100, 200, 400 and 800~mK.}
\end{figure}

The conditions for minimizing  $T_N$  are more evident when
Eq.~(\ref{eq:TN}) is written as
\begin{equation}
T_N =\frac{\varepsilon\omega_0\zeta} {2k_B}\left[\frac{1}{Q\alpha_e^2\zeta}+Q\alpha_e^2\zeta\right].
\label{eq:four}
\end{equation}
Here $\alpha_e^2=M_{i}^2/(L_{T}L_s)$ is the effective coupling
constant between the SQUID and the input circuit. Minimizing $T_N$
involves setting $ Q\alpha_e^2\zeta= 1$, resulting in
\begin{align}
T_N^{\mathrm{(opt)}} =&\varepsilon\omega_0\zeta/{k_{B}} ,\quad
R_{\vphantom{N}}^{\mathrm{(opt)}}=
\alpha^2\zeta\omega_0L_{i},\nonumber\\
L_p^{\mathrm{(opt)}} =& (Q\alpha^2\zeta - 1)L_i, \label{eq:five}
\end{align}
where the coupling coefficient $\alpha^2=M_{i}^2/(L_{i}L_{s})$. For
our device $L_i = 1.1\,\mu$H, $M_i=7.1$~nH, and $L_s=80$~pH,
corresponding to $\alpha^2=0.58$.

For NMR at 1~MHz a $Q$ of the order of 100 is reasonable, then for
$\zeta = 1$ we obtain $L_p^{\mathrm{(opt)}} = 63\,\mu$H and
$R_{\vphantom{N}}^{\mathrm{(opt)}}=3.9\,\Omega$. We expected $\zeta
< 1$ so we wound a $47\,\mu$H pickup coil. By driving the pickup
coil via the transmitter we measured $Q=300$ at $\omega_0 = 2\pi
\times 884$~kHz, from which we inferred $R = 0.89\pm 0.03\,\Omega$
and $Q\alpha_e^2=3.9\pm 0.1$. Once $\zeta$ is known, full
optimization could be achieved by addition of a further resistive
element at coil temperature $T$. For these studies of noise as a
function of temperature this was not done, to eliminate potential
errors arising from temperature gradients.

\begin{figure}[t]
\includegraphics{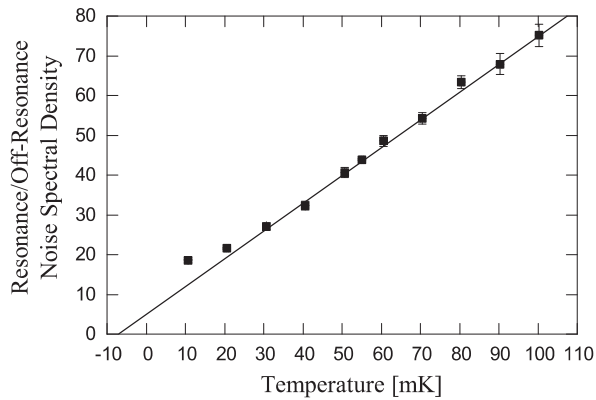}
\caption{\label{fig:noise_temperature} Resonance noise spectral
density normalised by the off-resonance noise for $T \le 100~$mK,
extrapolates to zero at $T = -T_{{N}}$ with $T_{{N}}=7\pm2~$mK.}
\end{figure}
\begin{table}[b]
\caption{Measured and predicted $T_N$ at five different working
points. Predicted $T_N$ obtained using Eq.~(\ref{eq:four}) for
$\zeta$ extracted from $r(0)$ values.}\label{tbl:results}
\centerline{\begin{tabular*}{86mm}{@{\extracolsep{\fill}} c c c c c c }
\hline
\hline
 Working  & $T_N$ [mK] & $T_N$ [mK]    & $\varepsilon\,[h]$ & $\zeta$           & $Q$       \\
 point    & measured   & predicted     &                    &                   &           \\
\hline
1 (open)  & $7\pm 2$   & $5 \pm 1$     & $26 \pm 1$         & $0.5 \pm 0.1$     & $300\pm1$ \\
2 (open)  & $8\pm 1$   & $5.4 \pm 0.6$ & $21 \pm 1$         & $0.7 \pm 0.1$     & $275\pm1$ \\
3 (FLL)   & $8\pm 2$   & $5.6 \pm 0.8$ & $24 \pm 1$         & $0.7 \pm 0.1$     & $250\pm4$ \\
4 (FLL)   & $9\pm 2$   & $6 \pm 1$     & $33 \pm 1$         & $0.6 \pm 0.1$     & $220\pm4$ \\
5 (FLL)   & $8\pm 2$   & $6 \pm 1$     & $24 \pm 1$         & $0.6
\pm 0.1$     & $270\pm5$\\
\hline
\end{tabular*}}
\end{table}

The temperature dependence of the noise in this tuned circuit was
measured at five different SQUID working points (two open loop and
three FLL), the results of which are summarized in
Table~\ref{tbl:results}. The output of the SQUID electronics was fed
to an HP 3588A spectrum analyzer with a negligible noise level. In
Fig.~\ref{fig:flux_noise} we show the frequency dependence of the
flux noise measured in open loop mode (working point 1) with the
pickup coil at 20~mK and the tuning capacitance at 10~mK. For a
high-$Q$ input circuit the off-resonance flux noise is equivalent to
that of an open circuit. We infer the open input flux noise $S_\Phi$
at $\omega_0$ from the off-resonance noise as shown by the dashed
line. This corresponds to $\varepsilon = 26\pm1~h$, where $h$ is
Planck's constant. The inset shows noise in the vicinity of
resonance as the pickup coil temperature is varied between 20 and
800~mK. The noise power at resonance, normalized by the
off-resonance noise power, can be written as
\begin{equation}
 r(T) = \frac{S_{\Phi}^{\mathrm{(res)}}} {S_{\Phi}} = 1 +
Q^2\alpha_{{e}}^{4}\zeta^2
+\frac{2k_{B}^{\phantom{1}}Q\alpha_e^2}{\omega_0\varepsilon}T.
\label{eq:six}
\end{equation}
This is plotted versus coil temperature in
Fig.~\ref{fig:noise_temperature}. These data are linear over a wide
temperature range. The data deviate from a linear fit at the lowest
temperatures possibly due to insufficient thermalization of the
pickup coil. We obtain $T_{N} = 7\pm2$~mK from a fit to the linear
region. We obtain an almost identical $T_N$ with the SQUID operated
in FLL mode (see Table~\ref{tbl:results}), a consequence of the low
parasitic coupling between feedback and input coils achieved through
careful SQUID design.

Within our model, at working point 1, we estimate $\zeta =
0.5\pm0.1$ from the value of $r(0)= 5\pm1$, obtained from the fit in
Fig.~\ref{fig:noise_temperature}. Eq.~(\ref{eq:four}) gives a
predicted $T_N$ of 14~mK for the na\"{\i}ve assumption of $\zeta =
1$, and $5\pm1$~mK using the measured value of $\zeta$. The slope of
the fit in Fig.~\ref{fig:noise_temperature} is less then that
obtained using Eq.~(\ref{eq:six}). This could be explained if a
noiseless resistance, arising from coupling to the SQUID, accounted
for $\approx 35~\%$ of the total input resistance.

Under optimum conditions tuning the input circuit gives an improvement
in $S/N$ approaching a maximum of order $\sqrt{Q}/\alpha$ if one cools
$R$ to well below $T_N$. This can be very significant and becomes easier
to achieve as $\omega_0$ increases.

In summary we have designed and built a tuned NMR spectrometer based
on a two-stage dc SQUID coupled to a conventional wire wound pickup
coil to study superfluid $^{3}\text{He}$ confined in submicron thick
cavities. Open loop and FLL tests of this spectrometer with the
pickup coil in a superconducting shield have shown that
$T_N~<~10~$mK is achievable, close to the prediction based on the
measured energy sensitivity. Improvements in energy sensitivity are
possible through operating the SQUID at temperatures lower than
1.4~K.

This work was supported by EPSRC grant EP/C522877/1 and in part by
EPSRC ARF EP/E054129/1.

\end{document}